\documentclass{icrc2009}

\usepackage{graphicx}   
\usepackage[caption=false]{caption}    
\usepackage[font=footnotesize]{subfig} 
\usepackage{fixltx2e}
\usepackage{url}

\newcommand{\shorttitle}[1]%
{\markboth{Proceedings of the 31\MakeLowercase{$^{st}$} ICRC, {\L}\'{o}d\'{z} 2009}{#1} }
\newcommand{\etal}{\MakeLowercase{\textit{et al. }}} 

\begin{document}

\title{Ultra high energy cosmic rays: subluminal and superluminal shocks}

\author{\IEEEauthorblockN{Athina Meli\IEEEauthorrefmark{1},
			     Julia K. Becker\IEEEauthorrefmark{2},
                         John Quenby\IEEEauthorrefmark{3}}
                            \\
\IEEEauthorblockA{\IEEEauthorrefmark{1} Erlangen Center for Astroparticle Physics, Universit\"at Erlangen-N\"urnberg, Germany}
\IEEEauthorblockA{\IEEEauthorrefmark{2} Institutionen f\"or Fysik, G\"oteborgs Universitet, 41296 G\"oteborg, Sweden}
\IEEEauthorblockA{\IEEEauthorrefmark{3} High Energy Physics Group, Blackett Laboratory, Imperial College London, Prince 
Consort Road SW7 2BZ, UK}}

\shorttitle{Meli \etal Ultra high energy cosmic rays: subluminal and superluminal shocks}
\maketitle

\begin{abstract}

Diffusive shock acceleration is invoked to explain non-thermal particle acceleration in Supernova Remnants, 
Active Galactic Nuclei (AGN) Jets, Gamma ray Bursts (GRBs) and various large scale cosmic structures.
The importance of achieving the highest observed particle energies by such a mechanism in a given astrophysical 
situation is a recurring theme.
In this work, shock acceleration in relativistic shocks is discussed, mostly focusing on a numerical study concerning proton 
acceleration efficiency by subluminal and superluminal shocks, emphasising on the dependence of the scattering model,
bulk Lorentz factor and the angle between the magnetic field and the shock flow. 
We developed a diffuse cosmic ray model based on the study of different shock boost factors,
which shows that spectra from AGN fit current observations of ultra high energy cosmic rays,
above 5.7 $\times$ 10$^{10}$ GeV, much better than GRBs, indicating that AGN are the primary 
candidates to explain the UHECR flux. Recent Fermi observations of GRB090816c indicate very flat spectra which
are expected within our model predictions and support evidence that GRB particle spectra can be
flat, when the shock Lorentz factor is of order $\sim$ 1000.

\end{abstract}

\begin{IEEEkeywords}
 relativistic shocks, acceleration, AGN, GRBs, cosmic ray origin 
\end{IEEEkeywords}

\section{Introduction \label{introduction}}

Active Galactic Nuclei (AGN) and Gamma Ray Bursts (GRBs) seem
to be the two most promising source candidates for the production of charged ultra high energy cosmic rays (UHECRs).
Work in the late 1970s by a number of authors, e.g. \cite{krymskii77,bell78a,bell78b}, basing their 
ideas on the original Fermi acceleration mechanism, i.e. ~\cite{fermi49, fermi54},
established the basic mechanism of particle diffusive acceleration in 
non-relativistic shocks.
Since then, considerable analytical and 
numerical investigations have been performed, but questions remain concerning details of the 
acceleration mechanism at highly relativistic shock speeds and the consequent production and origin 
of UHECRs.

\section{The microphysics of shocks - numerical approach}
In this two-fold work we firstly, present a series of Monte Carlo simulations with the aim to provide a more
refined determination of the possible accelerated particle shock spectra that can result and secondly, we develop a
diffuse cosmic ray (CR) model to explain the UHECR observed spectrum.
A Monte Carlo numerical approach solves the well known Boltzmann transport equation which depends on the 
assumptions that the collisions represent diffusive scattering in pitch angle and that the scattering 
is elastic in the fluid frame where there is no residual electric field, taking into account the theoretical result 
that Alfv{\'e}n waves are limited to $V_{A} \rightarrow c/\sqrt{3}$.
Then, the first order Fermi (diffusive) acceleration mechanism is simulated by following the test particles' 
guiding centres and allowing for numerous pitch angle scatterings in interaction with the assumed magnetised media,
while at each shock crossing the particles gain an amount of energy determined by a Lorentz transformation of 
a reference frame.
Standard theory assumes conservation of the first adiabatic invariant
in the so called de Hoffmann-Teller (HT) frame in order to determine reflection or transmission 
of the particles. Trajectory integration calculations, see \cite{hudson1965}, \cite{Parker65},
giving the phase dependence of the probability of transmission
as a function of phase and pitch angle, showed that the reflection percent plotted against 
pitch angle, never varied more than $20 \%$ from the mean value. 
In the relativistic shock situation, anisotropy renders the input to the shock from upstream, 
very anisotropic in pitch angle, but as discussed in \cite{MeliThesis}, it is an
acceptable approximation to randomise phase before transforming to the HT frame
and then to use the adiabatic invariant to decide on reflection or transmission.
Therefore, for our simulations the initial injection energy ($\gamma$) of the particles 
is taken to be $\gamma\sim(\Gamma+10)$, where $\Gamma$ is the shock Lorentz boost factor.\\
For the oblique shock cases, especially studied here (for a detailed description of the shock 
kinematics see \cite{Melietal08}), 
we apply a pitch angle  scatter [$1/\Gamma \leq \delta \theta \leq 10/\Gamma , \phi \in  (0, 2\pi)]$  which approximately
corresponds to a situation with a power spectrum of scattering waves, $P(k)/\,B^{2}=5/4 \sqrt{2} \cdot \Gamma^{-2}\cdot k^{-1}$ 
and with the neglect of cross-field diffusion, right up to the shock interface. 
Provided the field directions encountered are reasonably isotropic in the shock frame, we know that
 $\tan\psi_{1}=\Gamma_{1}^{-1}\tan\psi_{NSH}\sim \Gamma_{1}^{-1}\sim \psi_{1}$
where '1' and 'NSH' refer to the upstream and normal shock frames respectively and $\psi$ is the
 angle between the magnetic field and the shock normal.
The concentration of field vectors close to the $x$-axis in the upstream fluid frame allows 
a reasonable probability of finding a HT frame with a boost along the negative $y$-axis less than $c$.
Making this boost then yields an upstream HT frame inclination,
$\tan\psi_{HT,1}=\Gamma_{HT,1}\tan\psi_{1}$. While all particles are allowed to cross from downstream to upstream,
only particles with a critical HT frame pitch angle, $\theta_{c}$, given by $\theta_{c}=\arcsin (\frac{B_{HT,1}}{B_{HT,2}})^{0.5}$
are allowed to cross upstream to downstream and conservation of the first adiabatic invariant is used to determine the new,
downstream pitch angle ~\cite{newman1992}. The pitch angle is measured in the local fluid frame, while the value $x_i$ gives the distance of the particles to the shock front, where the shock is assumed to be placed at $x=0$. 
A compression ratio of 3 is used and although some MHD conditions favour a value of 4, the
study of \cite{meli_quenby03_2} does not find a substantial difference between the simulation results for 
these two cases. Away from the shock, the guiding centre approximation is used so that
a test particle moving a distance, $d$, along a field line at $\psi$ to the shock normal,
in the plasma frame has a probability of collision within $d$ given by $P(d)=1-\exp(-d/\lambda)=R$,
where the random number $R$ is $0 \leq R \leq 1$. Weighting the probability by the current in the 
field direction $\mu$ (i.e. $\cos\theta$) yields $d=-\lambda \mu \ln R$. \\
For a detailed description on the numerical method see \cite{Melietal08}.

\section{Results - Discussion \label{results}}

\subsubsection{Superluminal shock spectra}

We follow the helical trajectory of the particle until it intersects
the shock front (applying pitch angle  scatter [$1/\Gamma \leq \delta \theta \leq 10/\Gamma , \phi \in  (0, 2\pi)]$ 
as mentioned above).
Simulation runs demonstrated that the results were almost independent of $\psi$.  
Shock frame spectra are illustrated in figure 1 with a simulation run for an arbitrarily chosen, typically large
inclination angle, $\psi=76^{o}$, employing a range of shock boost factors.
We see that superluminal relativistic shocks are not efficient accelerators for very high energy particles 
and are unlikely to contribute to observable effects as we will see lateron. These conclusions concur with the work
of \cite{niemec_ostrowski_icrc07}.

\begin{figure}[t]
\centering
\includegraphics [width=7.0cm]{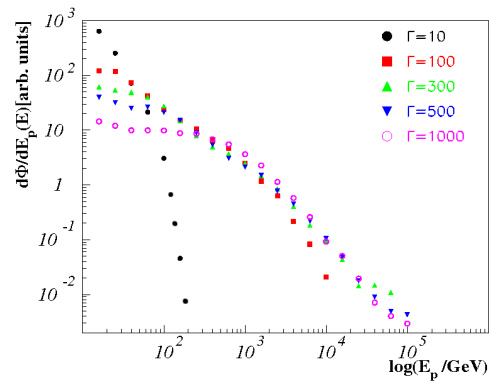}
\caption[Superluminal, relativistic spectra at $\psi=76^{\circ}$]{Superluminal, 
relativistic spectra at $\psi=76\deg$. Shock boost factors are varied between 
$\Gamma=10,100,300,500,1000$. Spectra for different inclination angles $\psi$ are comparable. 
In the region of efficient acceleration, the spectra approximately follow
power-laws with spectral indices lying between $\sim 2.0 - 2.3$. }
\label{super_spectra:fig}

\end{figure}

\begin{figure}[t]
\centering
\includegraphics [width=7cm]{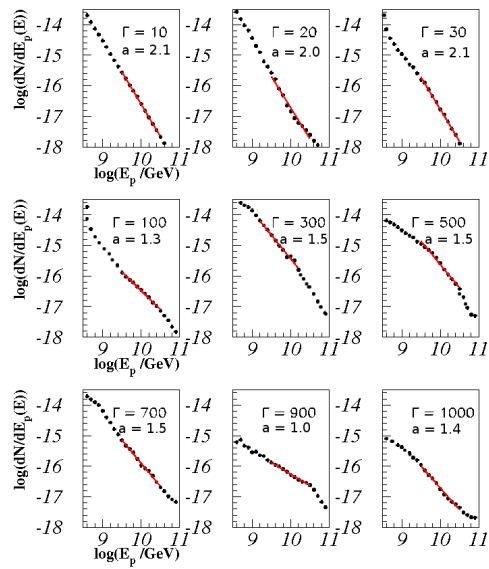}
\caption{Subluminal spectra averaged over three shock angles ($\psi=23^{\circ},\,33^{\circ}$ and $43^{\circ}$)
for different $\Gamma$: $\Gamma=10,20,30$ is displayed in the first row, in the middle,
  $\Gamma=100,300,500$ is shown and $\Gamma=700,900,1000$ is the bottom
  row. The black circles in each graph represent the simulation result. The red (solid)
  lines show the single power-law for comparison and the spectral indexes (a) values are shown accordingly. \label{mean_all}}
\end{figure}

\begin{figure}[h]
\centering
\includegraphics [width=7.5cm, height=7cm]{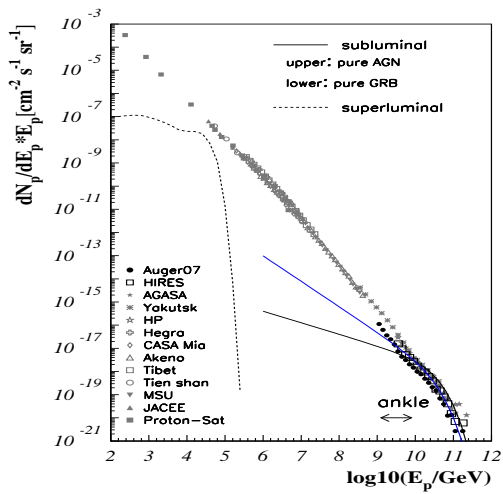}
\caption[Diffuse primary flux from GRBs and AGN]{The maximum predicted diffuse flux from GRBs and AGN with superluminal shock fronts (dashed line) and subluminal shocks (solid blue (upper) and black (lower) lines). For subluminal
sources, the upper line is a pure AGN-produced spectrum, the lower line represents 
a pure GRB spectrum. The flux is compared to the measured CR
spectrum. In the case of superluminal sources, 
50\% is assumed to come each from GRB and  50\% from AGN.}
\label{sl_cr}
\end{figure}

\begin{figure}
\centering
\includegraphics[width=6.5cm,height=5cm]{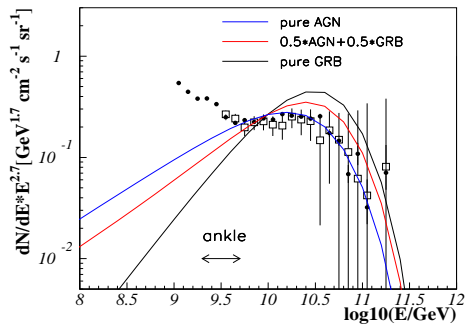} 
\caption[Spectrum of UHECRs multiplied by $E^{2.7}$.]{Spectrum of UHECRs
  multiplied by $E^{2.7}$. Data points from Auger,
\cite{yamamoto_icrc2007} and HiRes, \cite{hires_spect2002}. The solid lines
represent the same predictions as presented in Fig.~\ref{sl_cr}. Auger data
have been renormalised at $10^{10.3}$~GeV to HiRes data and the calculated
spectra have also been normalised to HiRes data. The data can
be described well by a pure AGN spectrum (blue (lower) line) within experimental
uncertainties and test particle acceleration accuracy. The red (middle) line is a
mixture of 50\% GRBs contribution and 50\% AGN contents, the black (upper) line is a
pure GRB spectrum.}
\label{sl_cr_e27}
\end{figure}

\subsubsection{Subluminal shock spectra}
Particle spectra produced in relativistic subluminal shocks with the numerical method of \cite{Melietal08} 
have been calculated for three different inclination angles in the shock frame,
$\psi=23^{\circ},\,33^{\circ}$ and $43^{\circ}$. Shown here (figure 2) are particle spectra 
averaged over the three inclination angles at a particular $\Gamma$ (where $\Gamma=10,100,300,500,1000$). 
The averaged values give a more realistic estimate of the diffuse particle flux (to be discussed later) from extragalactic sources, 
since a range of angles is more likely to occur in the AGN and GRB shocks. A power-law fit to the simulated spectra
between $10^{9.5}$~GeV and $10^{10.5}$~GeV is taken.
We concentrate on the highest energies because observation of particle-induced air showers indicate that at
energies between  $10^{9.5}$~GeV and $10^{10.5}$~GeV, the origin of the charged CR flux is extragalactic
and that this flux is distinct from a dominant galaxy produced component at lower energies. 
At even higher energies, the spectra will be modified in practise
by the absorption of protons due to interactions with the cosmic microwave
background. \\
Regarding relativistic shocks and their characteristics, \cite{Steckeretal07} investigating parallel shocks up 
to $\Gamma=30$ found an increase in structure in the spectral
shape and a decrease in slope as $\Gamma$ increased with a dependence $E^{-1.26}$ at $\Gamma=30$. These
trends are found in our work, extending to far higher $\Gamma$ factors and for a more general set of
subluminal shock inclination angles. Furthermore, \cite{bednarz_ostrowski98} employed pitch angle scattering and 
varying cross-field diffusion coefficients and found that at low $\Gamma$, steep spectra occurred at large inclination 
angles but all values of these parameters seemed to produce spectral slopes of -2.2 at $ \Gamma=243$. In contrast   
\cite{meli_quenby03_1} found spectra flatter than $E^{-2}$ for parallel shocks as $\Gamma\rightarrow1000$.
In addition, related work by \cite{Niemec05} with wave spectra 
$P(k)\sim k^{-1\rightarrow1.5}$, found spectra flatter than $E^{-2}$ with 
noticeable spectral structure in inclined, subluminal shocks at upstream velocities of 0.5c and a weakly perturbed field.
Their trajectory integrations took into account cross field diffusion. 
There is an agreement that at very high inclinations, significant acceleration above 
that due to a single shock cycle is ruled out. \\
One realises that there seems to be a developing consensus that high $\Gamma$
subluminal shocks result in flatter spectra than $E^{-2}$. 
In \cite{Dingus95}, it is shown that GRBs
can have flat relativistic electron spectra.  
A number of varying power-law slopes are also consistent with radio data on
the electron spectra injected at terminal hotspots in the lobes of  
powerful FR-II radio galaxies  i.e. \cite{Machalskietal07} among others. 
Recenlty, Fermi observations of GRB090816c \cite{Fermi09} indicate very flat spectra,
to be discussed lateron.

\subsubsection{The diffuse spectrum of CRs\label{spectratodiffuse}}
The source proton spectra derived above can be translated into an
expected diffuse proton flux from astrophysical sources by folding
the spectra with the spatial distribution of the sources. 
In our calculations, adiabatic energy losses are taken into account and it is  also assumed that both AGN and GRBs
follow the star formation rate  to determine the
number density evolution with comoving volume, see for
  example, \cite{hasinger05} in the case of AGN and \cite{pugliese00} in the
  case of GRBs. Also the absorption of protons due to the interactions with the CMBR at
  $E_p >5\cdot 10^{19}$~eV, as was recently confirmed by the Auger experiment, see \cite{yamamoto_icrc2007},
is taken into account. Furthermore, because the calculated
particle spectra are given in arbitrary units, normalisation of the
overall spectrum as measured at Earth is achieved using observation. \\
(i) In the case of superluminal sources,  normalisation of the expected signal follows from the most
restrictive upper limit on the neutrino signal from extraterrestrial sources
given by the AMANDA experiment (\cite{jess_diffuse}),
$ E_{\nu}^{2}\,\frac{dN_{\nu}}{dE_{\nu}}<7.4\cdot 10^{-8}\frac{\mbox{GeV}}{\mbox{s sr cm}^2}$ .\\
(ii) In the case of subluminal sources, using neutrino flux limits leads
to an excess above the observed spectrum of charged CRs, since the limits are
not stringent enough yet. Instead, the measured CR spectrum
above the 'ankle' is used to estimate the
contribution from subluminal sources. The CR energy flux above the
ankle is given by e.g. \cite{wb97}.\\
It is expected that the contribution comes from a combined signal from AGN
and GRBs. It is assumed that the fraction of UHECRs coming
from AGN, contributes a fraction $0<x<1$. Therefore, the fraction of UHECRs from GRBs is $(1-x)$.
Thus, the total spectrum as observed at Earth is given as:\\
$
\frac{dN_p}{dE_p} = A_p \int_{z_{\min}}^{z_{\max}} (  ( x \cdot 
$
$
\frac{d\Phi_{AGN}}{d E_p(z)}(E_p(z))+(1-x) \cdot \frac{d\Phi_{GRB}}{d E_p(z)}(E_p(z))  )
\cdot \exp\left(-\frac{E_p(z)}{E_{cut}(z)}\right)\cdot g(z)dz \, .
$

The minimum redshift is set to $z=0.001$. This excludes only the closest
AGNs which contribute less than 1\% according to newest
results of \cite{auger07}, \cite{becker_biermann08}.
The maximum redshift is taken to be
$z_{\max}=7$. As the main contribution comes from redshifts of $z\sim 1-2$ due
to the high number of sources at these redshifts, the exact values of the
integration limits are not crucial.\\
The diffuse spectrum from superluminal and subluminal shock sources (dashed, solid lines) as could be
measured at the Earth is shown in figure 3 and the diffuse spectrum multiplied by  $E^{2.7}$, 
focusing on energies $> 10^{9}$ GeV for subluminal shock sources,  is shown in figure 4. 
One sees a very good model agreement (blue (lower) line of figure 4).

As aforementioned, Fermi observed the GRB090816c with both instruments,
GBM and LAT \cite{Fermi09}. This means that the spectral behaviour of this burst is
studied between 10 keV and 10 GeV, i.e. over 6 orders of magnitude. The
observed spectrum is believed to arise from electron synchrotron
radiation. The observation of the highest photons gives a limit on the
boost factor, as no cutoff due to pair production was observed. The
minimum boost factor is varying for the different time bins and it ranges
between $\Gamma_{\min} \sim 600 - 1100$. The photon spectrum at energies
below the break energy, where it is expected to be unaffected by
absorption effects, shows a spectral behavior between $\nu^{-0.6} -
\nu^{-1}$. In the fast cooling regime, this implies primary electron
spectra of $E^{-1.2} - E^{-2.0}$. These very flat spectra are expected
within our model and supporting evidence that GRB particle spectra can be
flat. Note however, that electrons easily interact
and primary spectra are therefore easily modified. Hence, it is not clear
yet if the observed spectrum is indeed unmodified. The observation of
neutrino spectra from GRBs can help to solve that matter, as neutrinos
freely escape from the source, see e.g. \cite{becker08}.

\section{Summary \& conclusions\label{summary}}
We have presented Monte Carlo simulation studies of the acceleration 
of test particles in relativistic, subluminal and superluminal shock environments (i.e. AGN, GRBs)
and the resulting CR spectra were used to calculate a diffuse CR contribution. 
Our results can be summarised as follows:\\
\underline{1. Subluminal} shock acceleration was studied with a pitch scattering angle.
The resulting spectral slopes were roughly independent of inclination angle, though some details of the features 
were different. A dependence of the spectral index $\alpha_p$ on the shock boost factor $\Gamma$ 
was found, leading to spectra of $\alpha_p\sim 2.0 - 2.1$ for mildly-relativistic shocks of
$\Gamma\sim 10-30$, but producing much harder spectra ($1.0<\alpha_p<1.5$) for highly-relativistic
shocks, $100<\Gamma<1000$.
The particle spectra arising from relativistic shocks in GRB
with very high boost factors between $100<\Gamma<1000$, have spectra flatter than
${E_{p}}^{-1.5}$.
The above findings are supported by the work at lower $\Gamma$ of \cite{Niemec05} and
\cite{Steckeretal07}. Observational evidence,
see \cite{Dingus95}, regarding irregular and flat spectra from GRBs may be explained by the spectra we present.\\
\underline{2. Superluminal} shocks, applying the same pitch angle scatter as above, 
seem only efficient in accelerating CRs up to $E_p \sim 10^{5}$~GeV, resulting in spectral indices of  
$\alpha_p\sim 2.0-2.3$. On the other hand, subluminal shocks are more efficient and able to accelerate 
CRs up to $E_p \sim 10^{12}$~GeV, factors
of $10^{9\rightarrow11}$ above the particle injection energy. 
Extragalactic, superluminal shocks energy density at low energies may be quite high
compared to the observed flux of UHECRs, but hidden by galactic
cosmic rays. Nevertheless they could be good candidates for  the production of 
high energy neutrinos and photons, see \cite{Melietal08}.\\
\underline{3.  Diffuse UHECR flux contributions} of AGN and GRBs was discussed.
For superluminal sources, such contributions can
be excluded by using current neutrino flux limits to normalise the spectrum.
In the case of subluminal sources, the
spectrum is normalised to the CR flux above the ankle,
$E_{\min}=10^{9.5}$~GeV. By applying AGN (with shock boost factors of $\Gamma=10$) as potential 
UHECRs sources, the spectrum fits very well the current CR data within experimental uncertainties. 

Recent Fermi observations of GRB090816c indicate very flat spectra which
are expected within our model predictions and support evidence that GRB particle spectra can be
flat, when the shock Lorentz factor is of order $\sim$ 1000.

\end{document}